\documentclass[electronic]{vgtc}             

\ifpdf
  \pdfoutput=1\relax                   
  \usepackage{graphicx}                
  \DeclareGraphicsExtensions{.pdf,.png,.jpg,.jpeg} 
\else
  \usepackage{graphicx}                
  \DeclareGraphicsExtensions{.eps}     
\fi%

\graphicspath{{figures/}{pictures/}{images/}{./}}

\usepackage{microtype}                 
\PassOptionsToPackage{warn}{textcomp}  
\usepackage{textcomp}                  
\usepackage{mathptmx}                  
\usepackage{times}                     
\usepackage{cite}                      
\usepackage{tabu}                      
\usepackage{booktabs}                  
\usepackage[caption=false, font=footnotesize]{subfig}

\onlineid{0}

\vgtccategory{Research}

\vgtcinsertpkg

\title{VR Research at Fraunhofer IGD, Darmstadt, Germany}

\author{
	Wolfgang Felger\thanks{felger@web.de}\\ %
     \parbox{1.4in}{\scriptsize \centering Retired, Germany} %
\and Martin Göbel\thanks{martin.goebel@email.de}\\ %
     \parbox{1.6in}{\scriptsize \centering Hochschule Bonn-Rhein-Sieg,\\ Germany} %
\and Dirk Reiners\thanks{dirk.reiners@gmail.com}\\ %
	 \parbox{1.4in}{\scriptsize \centering University of Central Florida, USA} %
\and Gabriel Zachmann\thanks{zach@cs.uni-bremen.de}\\ %
	 \parbox{1.4in}{\scriptsize \centering University of Bremen,\\ Germany}
}

\abstract{
	We present a historical outline of the research and developments
	of Virtual Reality at the Fraunhofer Institute for Computer Graphics (IGD)
	in Darmstadt, Germany, from 1990 through 2000.
} 

\CCScatlist{
  \CCScatTwelve{Human-centered computing}{Virtual REality}{History of VR}{Fraunhofer IGD};
}

\teaser{
 \centering
 \subfloat{%
	\includegraphics[width=0.24\linewidth]{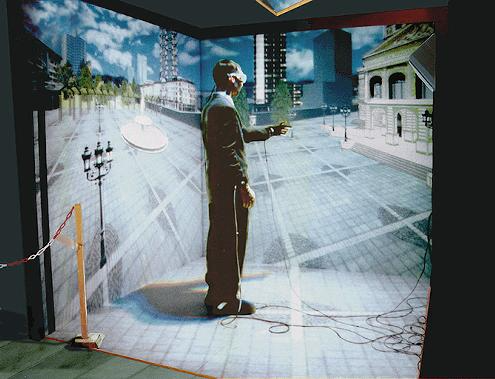}
 }\hfill
 \subfloat{%
	\includegraphics[width=0.24\linewidth]{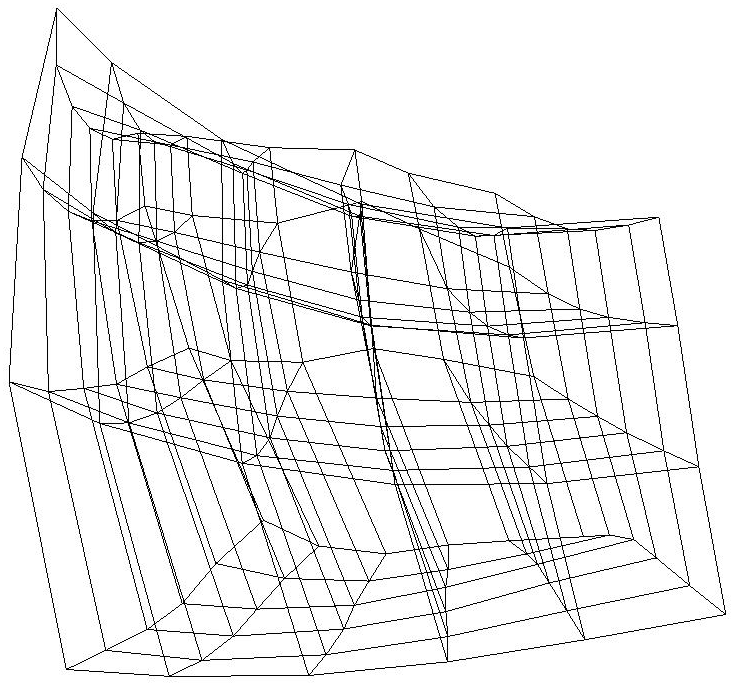}
 }\hfill
 \subfloat{%
	\includegraphics[width=0.24\linewidth]{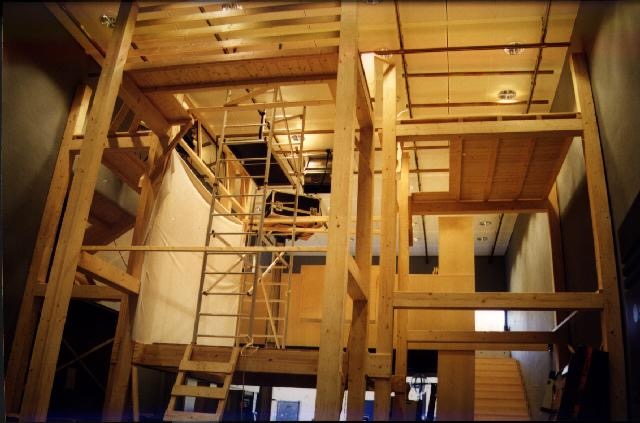}
 }\hfill
 \subfloat{%
	\includegraphics[width=0.24\linewidth]{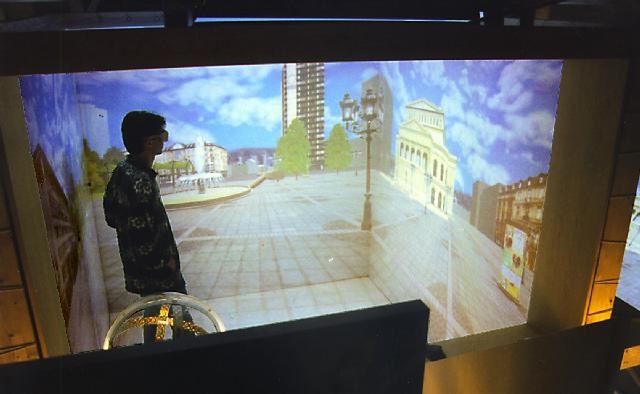}
 }
	\caption{Caves at Fraunhofer IGD:
				a) three-wall Cave at Fraunhofer IGD in 1993. 
				b) Distortion of the electro-magnetic tracking
				we measured in the Cave so that we could correct it,
				in order to have a good match between the real
				and the virtual viewpoint~\cite{Zachmann-1997-DistortionCorrection}.
				c) construction of the 5-walled cave.
				d) The 5-wall Cave in operation.
			}
	\label{fig:cave}
}

\begin{document}

\maketitle

\section{Introduction}

In this article, we will present and describe some of the developments
of Virtual Reality (VR) at the Fraunhofer Institute for Computer Graphics,
more precisely its department for visualization and simulation (A4),
later to be renamed into department for visualization and virtual reality%
\footnote{
	In the following, we will use the short term ``Fraunhofer'' or ``IGD''
	whenever we are talking of this department of the Fraunhofer IGD
	research institute.
}
in Darmstadt, Germany, from 1991 until 2000.

Research into VR started at this department in 1991 with funding by the German government for
a project called ``Vis-{\`a}-Vis''~\cite{Astheimer1992}, which allowed the department
to acquire a Silicon Graphics 3000, a VPL dataglove, and a Flight Helmet.

Towards the end of 1992,
Larry Rosenblum visited Fraunhofer IGD, where he recommended to create a European workshop on VR.
This led to the first
Eurographics Workshop on Virtual Environments (EGVE), co-located with the Eurographics
conference, in Barcelona in 1993, 
and the second one, co-located with IMAGINA, in Monte Carlo, Monaco, in 1995~\cite{Goebel1995}.

The overall state-of-the-art of VR in Europe until 1994 is described in~\cite{Encarnacao1994}.

\section{VR Hardware at IGD}

\begin{figure*}
 \centering 
 \subfloat{%
	 \includegraphics[width=0.25\linewidth]{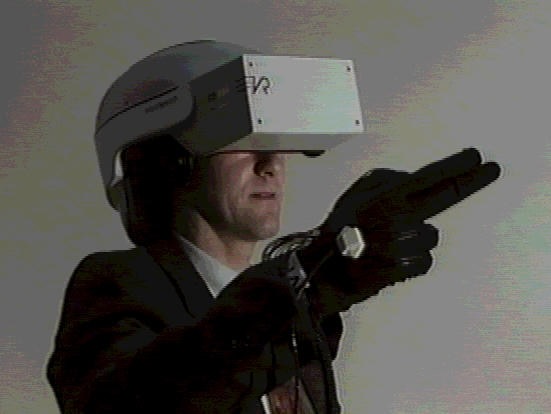}
 }\hfill
 \subfloat{%
	 \includegraphics[width=0.18\linewidth]{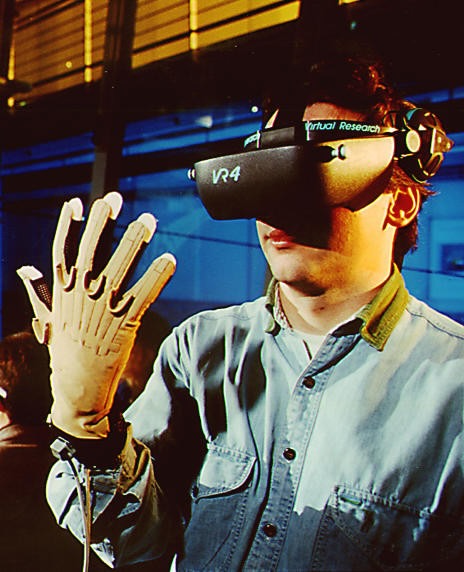}
 }\hfill
 \subfloat{%
	 \includegraphics[width=0.3\linewidth]{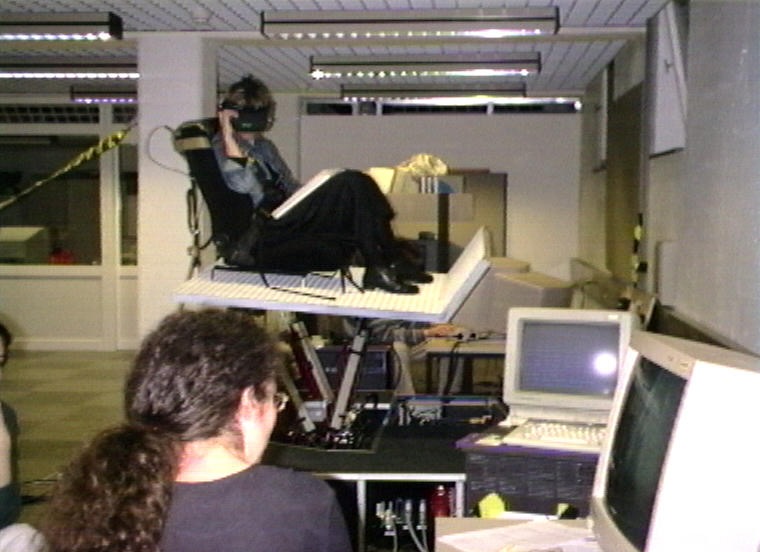}
 }
 \caption{Some of the VR devices used at Fraunhofer IGD, 
			roughly arranged from left to right according to their acquisition:
			a) VPL dataglove, Polhemus tracking sensor, and Virtual Research's Flight Helmet,
			b) Cyberglove and VR4 head-mounted display,
			c) Steward motion platform,
		}
 \label{fig:hardware}
\end{figure*}

The first VR devices at Fraunhofer was the legendary VPL dataglove
(see Figure~\ref{fig:hardware}).
In the beginning, it was tracked by a small Polhemus system (electro-magnetic tracking),
and rendering was done on an SGI 3000.
Later, the VPL glove was replaced by the standard Cyberglove, the Flight Helmet
was replaced by the VR4 model, and Polhemus' long-range ``disco ball'' transmitter
was used. This combination of devices was then used for several years, 
since it proved to be a very good compromise between price, usability, and technonological standards.

In addition, Fraunhofer IGD installed a show room for virtual reality in 1993,
which comprised a large stereo projection screen with a two-projector back projection
to be viewed using polarization glasses~\cite{Goebel1993}.
The stereo projection was driven by an SGI Crimson Reality Engine with multi-channel extension.
(More technical details can be found in~\cite{Felger93}.)
The show room provided space for about 20 viewers.
Demos were run by two presenters, one of them wearing the HMD and performing the demo, 
the other one explaining the technology and what was going on.
The idea was to have a facility that would allow potential customers to experience
the technology and its opportunities en masse.
Also, it greatly helped run demos on short notice, in particular, since the technology at the 
time was not very robust and failed to run very often for the slightest of reasons.

In the mid 90s, Fraunhofer IGD installed its first Cave
(Fig.~\ref{fig:cave}a).
It was a 3-wall Cave (2 walls, plus floor), with Polhemus tracking
and shutter glasses.
Due to heavy metal in the walls and ceiling of the room, we developed
a method to correct the distortion it induced in the electro-magnetic tracking
(Fig.~\ref{fig:cave}b).

In 1997, 
Fraunhofer IGD replaced the 3-wall Cave with a 5-wall Cave
(Fig.~\ref{fig:cave}, (c) and (d)).
In an effort to provide maximum rendering fidelity,
it was constructed inside a wooden frame that was housed in a huge room, 
such that the electro-magnetic field of the Polhemus tracking system would
get distorted as little as possible.
Each wall was driven by one projector, and one mirror per projector was used
to increase the throw distance from projector to wall.
Users had to wear shutter glasses (Crystal Eyes) and felt slippers over their shoes.
Rendering was done by one SGI rack Onyx. 





\section{The VR System Virtual Design}

\begin{figure}
 \centering 
	\includegraphics[width=0.49\columnwidth]{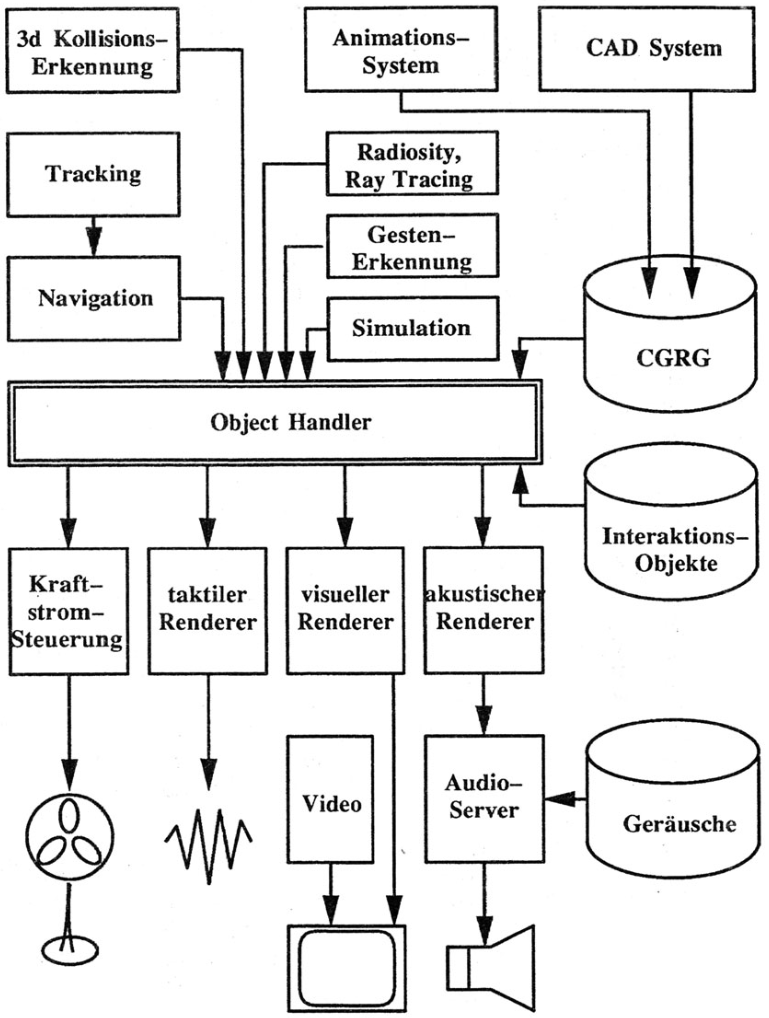}
	\includegraphics[width=0.49\columnwidth]{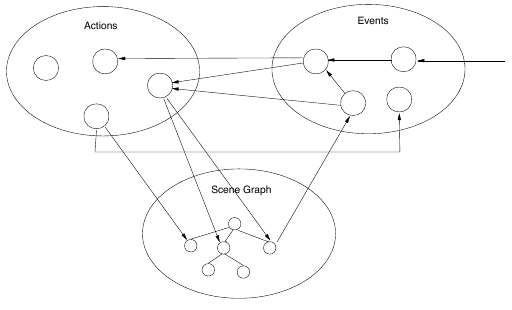}
	\caption{Software architecture of the VR system at Fraunhofer IGD,
			left: VD1, right: VD2. }
	\label{fig:vd-architecture}
\end{figure}

After the first input devices were mastered and first
experiments showed great promise, existing software at Fraunhofer IGD
for scientific visualization was forked off and built upon to 
become Fraunhofer's first VR software, called 
\textsl{Virtual Design} (VD1)~\cite{Astheimer1994,Astheimer1993}.
It was written in C and grew mostly on demand by some of the earliest 
demos and projects at Fraunhofer IGD.
Its software design already reflected the central role of
what would later be called a scene graph, but there was no scene hierarchy 
yet, and no concurrently running modules (see Figure~\ref{fig:vd-architecture}).

Virtual Design II (VD2):
after gaining some experience with VR system software,
VD2 was a complete redesign of the software archiecture, starting 
with coding from scratch in C++~\cite{Astheimer1995a}.
Its nucleus was implemented by Dirk Reiners and Gabriel Zachmann
during a research stay from March 1994 until October 1994 
at the Beckman Insitute/NCSA, Urbana-Champaign,
Illionis, USA.
(For which opportunity we are still extremely grateful.)
It was designed and implemented from the beginning as a multi-pipe, multi-process
renderer, based on modern OpenGL.      
Also, it contained a module for collision detection, since it was felt that
this is important for implementing higher-level interactions later on.
The algorithms worked directly on the meshes that were also used for 
rendering, in an effort to provide high-fidelity collision
detection~\cite{Zachmann-1997-RealTimeExact,Zachmann-1998-RapidCollisionDetection}
Also, this module was running concurrently to the rendering, since
collision detection is notorious for exceeding time budgets~\cite{Zach94b}.
Later, natural interaction using the virtual hand was implemented on top of that
(see Figure~\ref{fig:ems}),
albeit just with simple rubber-band, sticking, and DoF constraints
metaphors~\cite{Zachmann-2001-NaturalRobust}.
Since tracking at the time was done primarily using electro-magnetic systems
(at IGD we used Polhemus), we also needed to
solve the distortion problem of those 
systems~\cite{Zachmann-1997-DistortionCorrection}.
Finally, VD2 featured one of the first concepts of describing
the scene and the interactions possible in the scene 
by way of kind of a behavior graph 
(see Figure~\ref{fig:vd-architecture} right) involving notions such as 
\textsl{actions} and \textsl{events}~\cite{Zachmann-1996-LanguageDescribing}.


\begin{figure*}
 \centering 
 \subfloat{%
	 \includegraphics[width=0.3\linewidth]{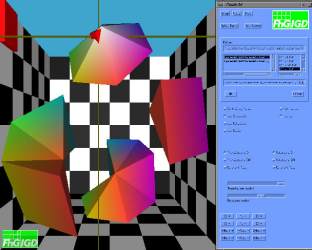}
 }\hfill
 \subfloat{%
	 \includegraphics[width=0.3\linewidth]{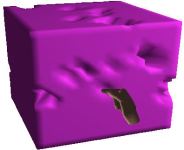}
 }\hfill
 \subfloat{%
	 \includegraphics[width=0.3\linewidth]{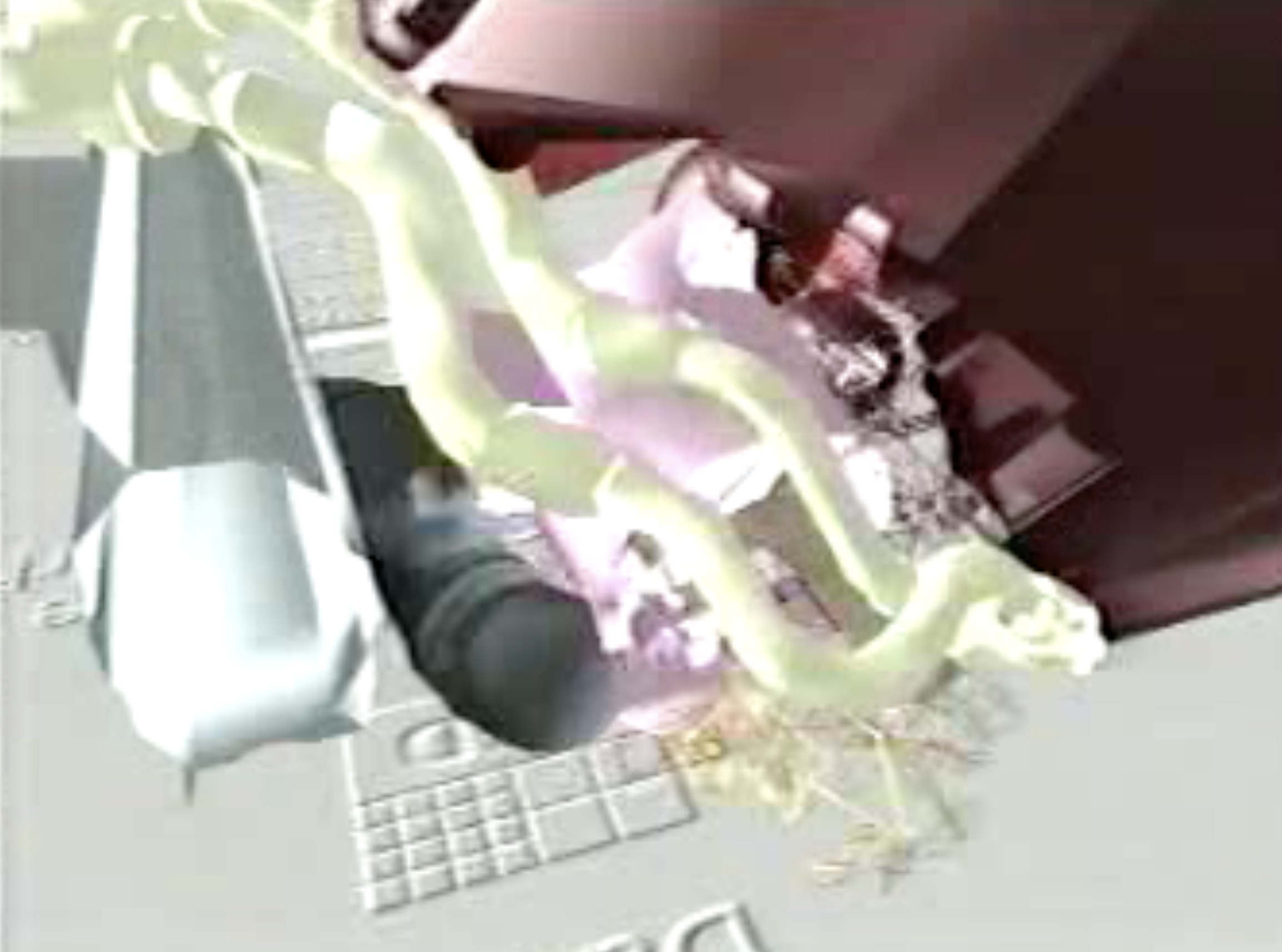}
 }
 \caption{Early applications and demos: (a) interactive 3D puzzle, (b) immersive freeform modelling,
		(c) AIT project, investigating the potential of VR for assembly simulation
		(obviously, some of the normals of the geometry are wrong).}
 \label{fig:early_apps}
\end{figure*}


%
%
%
%
%
%
%
%
%
%

\section{Early Applications}

One of the earliest applications at Fraunhofer IGD was kind of a 3D puzzle,
developed in 1991/92
(Fig.~\ref{fig:early_apps}a).
One of the goals was to investigate if VR devices (dataglove and spaceball) could
be more efficient than conventional input devices (mouse and dials)~\cite{Felger1992}.
At the time, applications in the scientific visualization area were envisioned 
for these kinds of devices, but, obviously, the results also apply to applications of VR,
such as assembly tasks in the automative and manufacturing industries.

Another early application involved very basic research into possibilities
of utilizing VR for sculpting 3D objects
(Fig.~\ref{fig:early_apps}b).

Finally, one of the early applictions tried to demonstrate the potential 
of VR for assembly simulation (see Figure~\ref{fig:early_apps}c).


\section{Demos and Shows}

The first demonstration of VR by Fraunhofer was done in 1992 at CeBIT, Hannover, Germany.
We showed the interactive applications 3D puzzle and free-form modeling
(see Fig.~\ref{fig:early_apps}) 
using the VPL dataglove.

\begin{figure*}
 \centering 
 \subfloat{%
	 \includegraphics[width=0.18\linewidth]{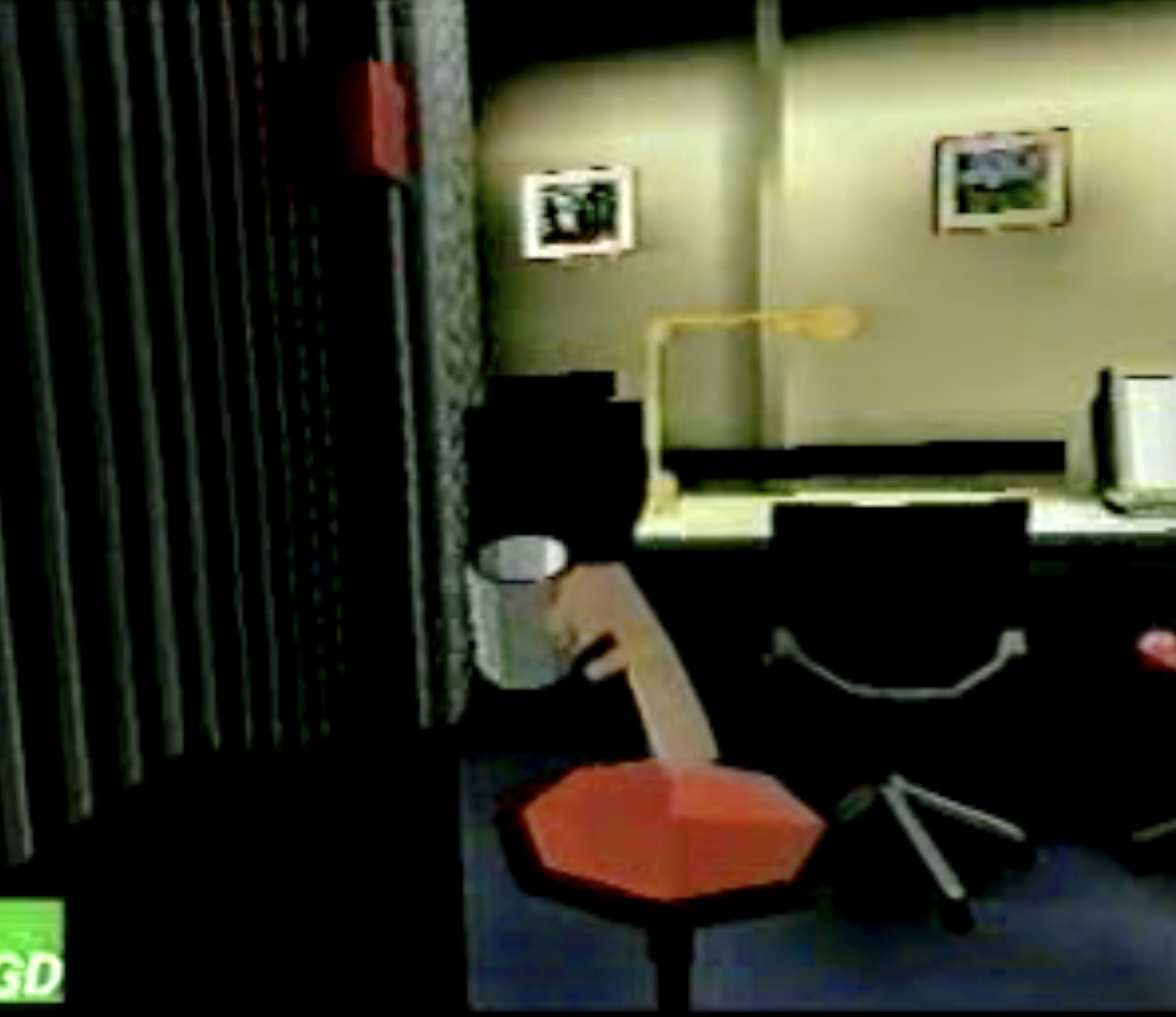}
 }\hfill
 \subfloat{%
	 \includegraphics[width=0.3\linewidth]{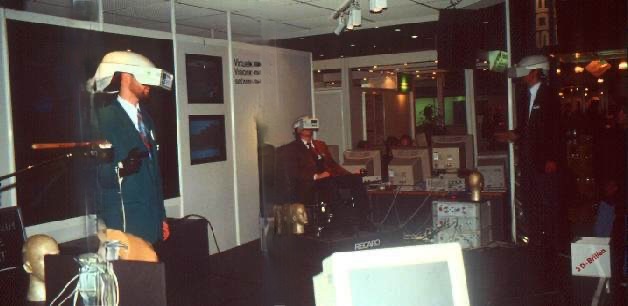}
 }\hfill
 \subfloat{%
	 \includegraphics[width=0.2\linewidth]{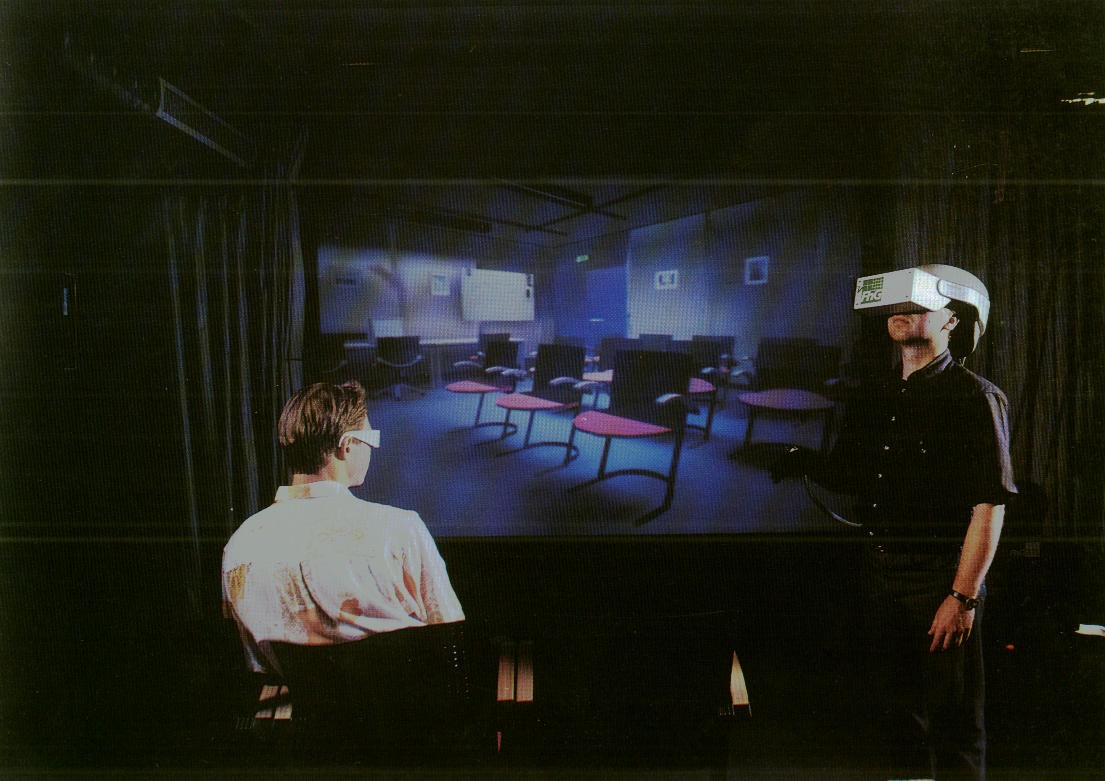}
 }\hfill
 \subfloat{%
	 \includegraphics[width=0.3\linewidth]{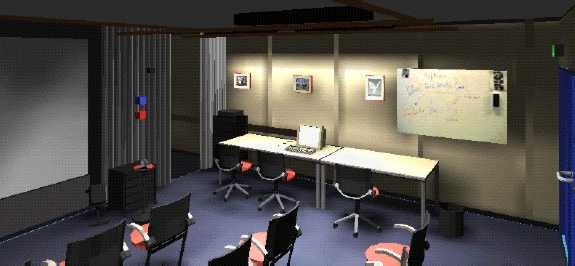}
 }
	\caption{Some of the early demos at Fraunhofer IGD:
			(a) first interactions (grasping and inverse kinematics),
			(b) CeBIT 1993,
			(c) the VR show room at Fraunhofer IGD,
			(d) virtual replica, demonstrated at Siggraph 1994 in a Cave.
			}
 \label{fig:demos2}
\end{figure*}

At CeBIT 1993,
we demonstrated a walk-through through an architectural design of a 
new airport in Abu Dhabi,
and some simple interactions such as grasping with the sticky metaphor
(see Figure~\ref{fig:demos2}, (a) and (b)).
The show attracted around 1000 visitors to the booth each day. 

At SIGGRAPH 1994, we presented a walk-through through the virtual replica
of the VR show room that was installed in 1993 at Fraunhofer IGD
(see Figure~\ref{fig:demos2} (a), (c), and (d)).
The demo was based on the first version of Virtual Design, with
precomputed lighting using IGD's radiosity software.
In a sense, the demo showed arguably the first instance of 
augmented virtuality, in that we placed a real cup in the real world
such that users would grasp it with their real hand when they 
were grasping the virtual cup with their virtual hand.

At the Hannover Industry Expo in 1995, we presented Virtual Reality
in Volkswagen's booth to both the public as well as Volkswagen's
executives. The demo involved animations of the engine, visualization
of a car body (Polo Harlekin), an interactive virtual wind tunnel
(see Figure~\ref{fig:demos2} (c)).
Later that same year, a precursor of the Detroit show (see below)
was demonstrated at the IAA auto show in Frankfurt.

\begin{figure}
 \centering 
 \subfloat{%
	 \includegraphics[width=0.52\columnwidth]{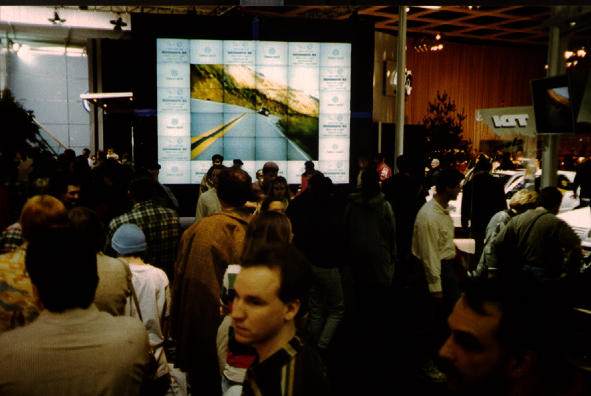}
 }\hfill
 \subfloat{%
	 \includegraphics[width=0.46\linewidth]{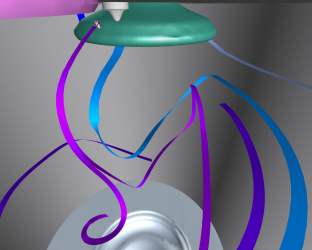}
 }
 \caption{Detroit auto show in January 1996 using VR to visualize the combustion processes
			in Volkswagen's diesel TDI engine.}
 \label{fig:detroit}
\end{figure}

By the mid 1990's, VR received considerable attention in the media and, thus,
appeared to be a rather exciting marketing vehicle.
Consequently, a relatively high-profile demo was developed together with Volkswagen 
for the Detroit auto show in January 1996.
It was presented in Volkswagen's booth once per hour for 4 weeks on end,
using an SGI Onyx and a passive stereo projection system for around 50
viewers. It has been a huge success in that every show was packed with long
waiting lines outside (see Figure~\ref{fig:detroit}a).
The demo involved a flight through the diesel engine, starting at the 
air inlets, through the combustion chamber, into the exhaust; 
in the combustion chamber, several phases of the whole combustion process were
visualized using standard scientific visualization techniques such as stream ribbons and
animated isosurfaces
(see Figure~\ref{fig:detroit}b).
This was 
performed and controlled live by a demonstrator, while the audience could watch in stereo
using polarization glasses.

\begin{figure}
 \centering 
	 \includegraphics[width=0.41\columnwidth]{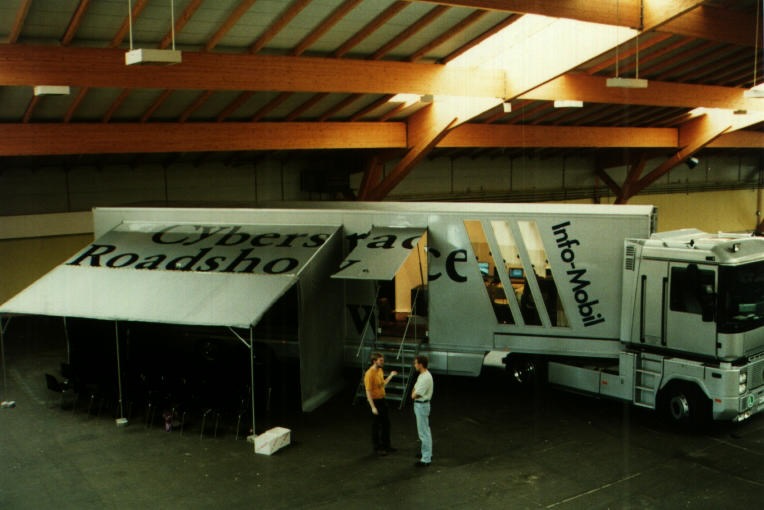} \hfill
	 \includegraphics[width=0.41\linewidth]{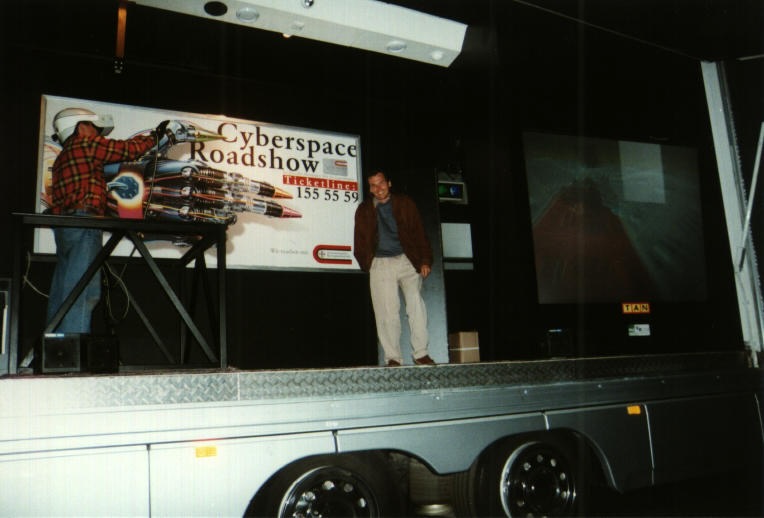} \hfill
	 \includegraphics[width=0.15\linewidth]{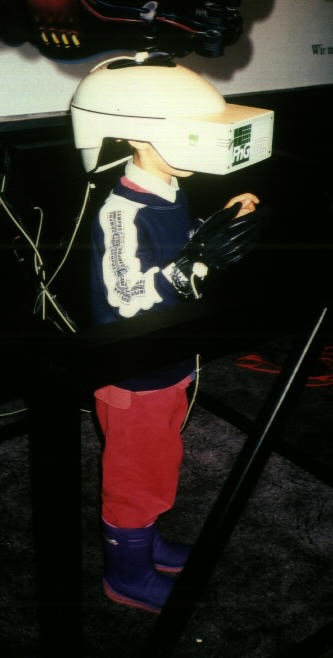}
 \caption{The marketing campaign for the UBS in Switzerland in 1994,
			using Fraunhofer's VR system and a VR show developed at Fraunhofer.}
 \label{fig:cybershow}
\end{figure}

Another marketing project was developed for the Swiss bank UBS to address specifically 
the young customer segment~\cite{Astheimer1995b}.
The marketing campaign was held during the summer 1994 with daily shows in 12 major Swiss cities.
The complete VR setup (SGI Crimson Reality Engine, passive stereo projection, HMD, tracking)
were installed in a truck that could be converted into kind of a showroom facility
(see Figure~\ref{fig:cybershow}).
Some of the lessons learnt regarded robustness of the VR setup, cost effectiveness,
but also diversity of the audience.

\section{Virtual Prototyping}

\begin{figure}
 \centering 
	\includegraphics[width=0.49\columnwidth]{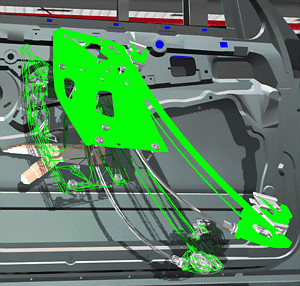}
	\includegraphics[width=0.49\columnwidth]{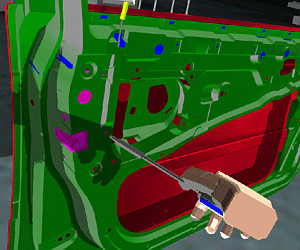}
	\caption{Examples of the research into VR as a tool for virtual prototyping, in collaboration with BMW.}
\label{fig:ems}
\end{figure}

One of the long-term research at Fraunhofer IGD was the investigation
of VR as a tool for virtual prototyping, in particular, assembly simulation in
VR~\cite{Dai-1996-VirtualPrototyping,Gomes-1998-IntegratingVirtual,Gomes-1999-VirtualReality}.
This involved questions such as manufacturability of new designs, involving rather
complex parts such as a car door~(see Figure~\ref{fig:ems}).
Users were able to grab parts and tools with their virtual hand (sticking metaphor),
use the tools such as a wrench 
(motion was then constrained by predefined constraints and inverse kinematics),
perform measurements (e.g., penetration depth), create live cross-sections, 
make recordings, and much more.

This project greatly spurred the further development of Fraunhofer's VR system
VD2. It showed that collision detection on the polygon-level was very important
to the engineers at BMW and inspired further research in that area
\cite{Zachmann-1998-RapidCollisionDetection}.

It also caused us to integrate real-time shadows (using OpenGL's shadow textures)
in the renderer, because it provided much better understanding of the spatial
relationships between the virtual obejcts for the users.

\begin{figure}
 \centering 
 \subfloat{%
	 \includegraphics[width=0.49\columnwidth]{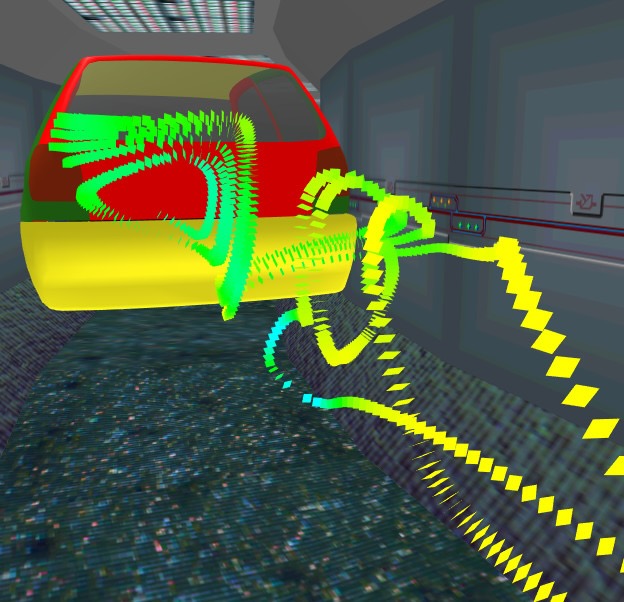}		
 }
 \subfloat{%
	 \includegraphics[width=0.49\columnwidth]{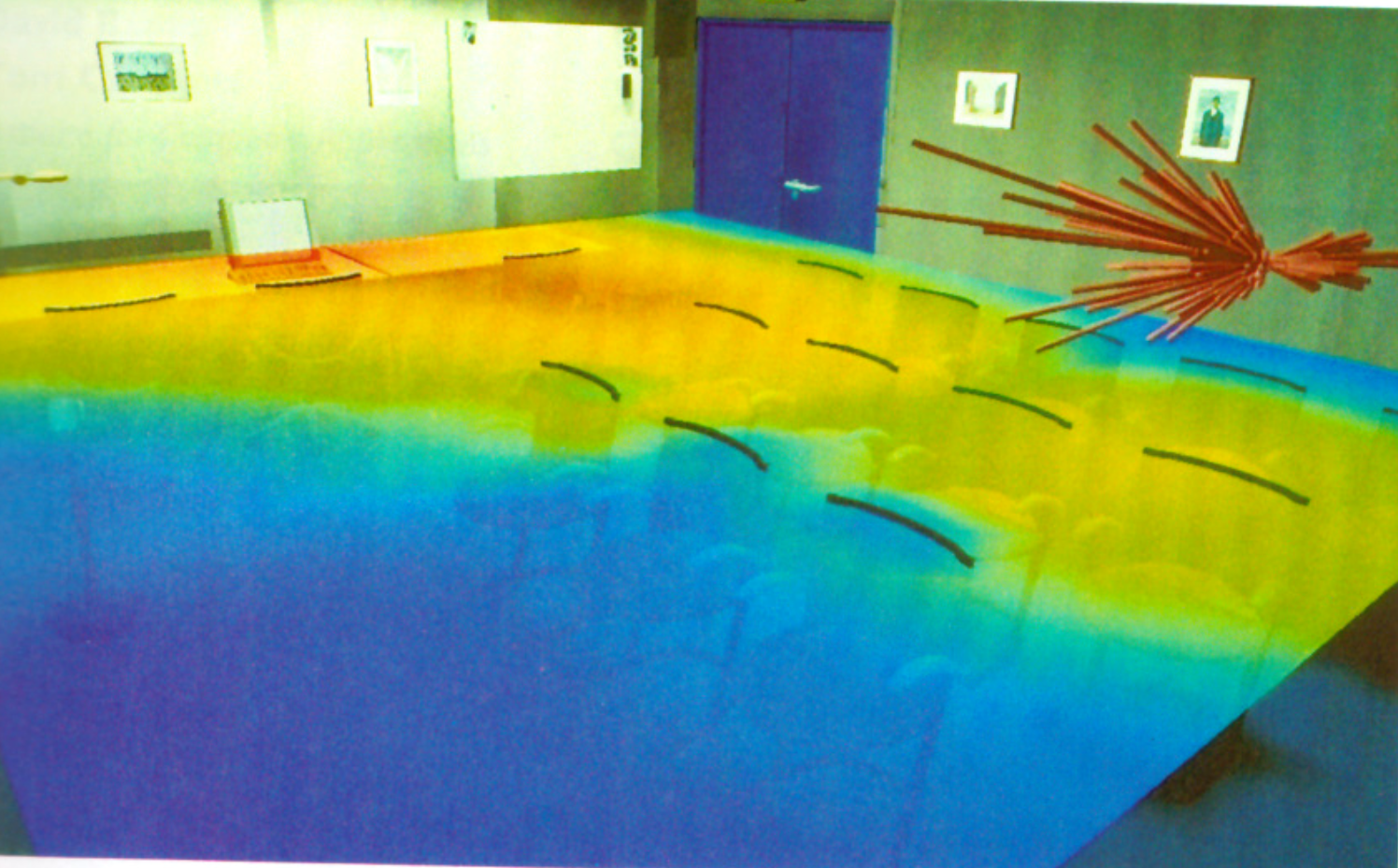}
 }
	\caption{Examples of immersive scientific visualization at Fraunhofer.
				Left: virtual wind tunnel. Right: global illumination and sound simulation results.
				}
	\label{fig:scivis}
\end{figure}

\section{Applications Demonstrating Feasibility of VR}

\begin{figure*}
 \centering 
	 \includegraphics[width=0.24\linewidth]{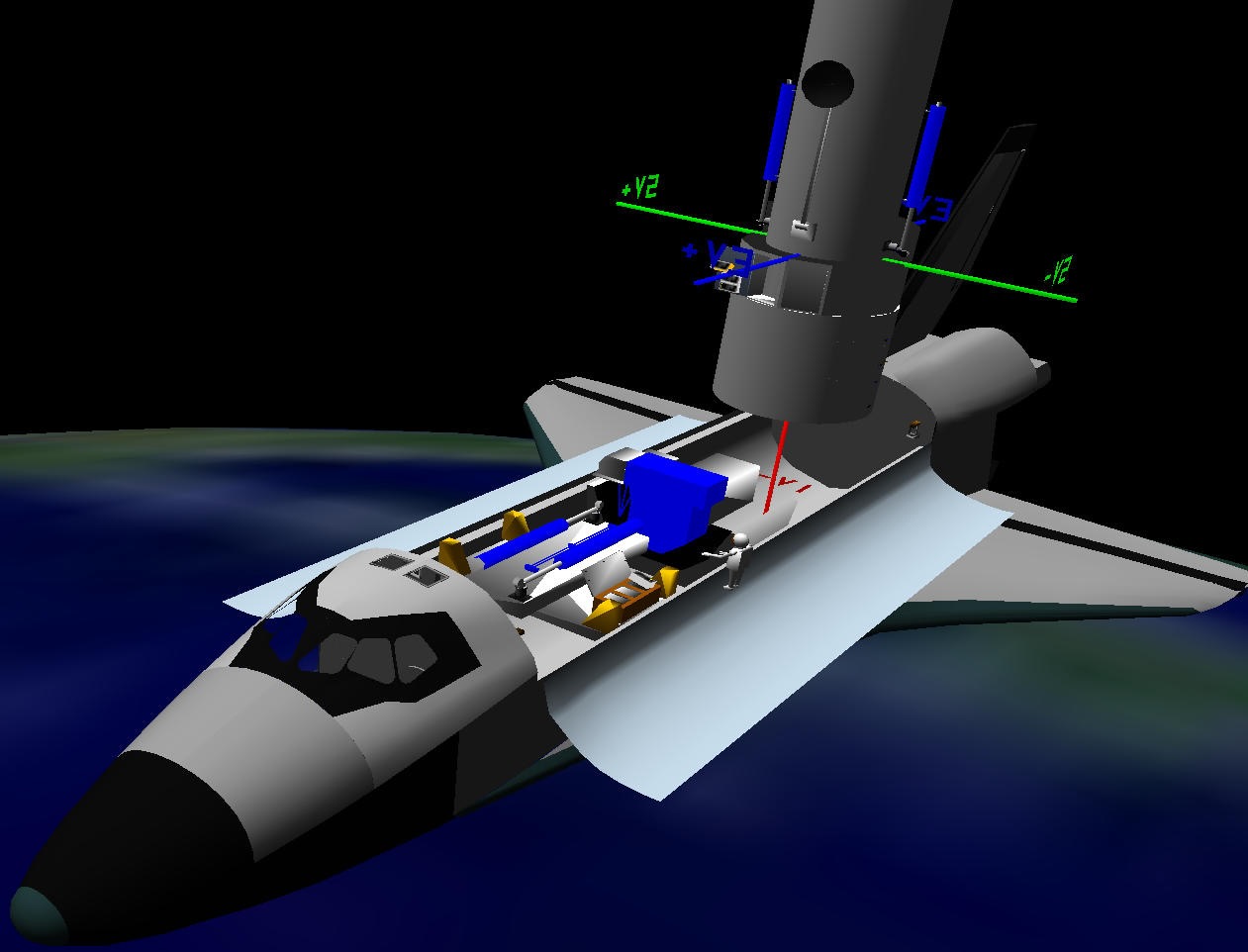} \hfill
	 \includegraphics[width=0.27\linewidth]{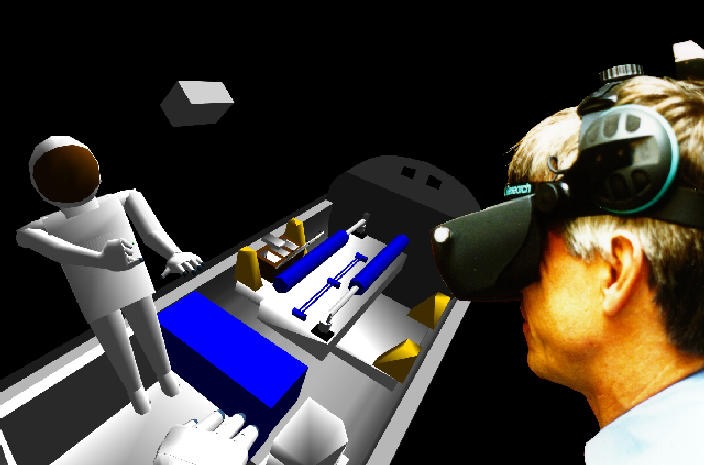} \hfill
	 \includegraphics[width=0.24\linewidth]{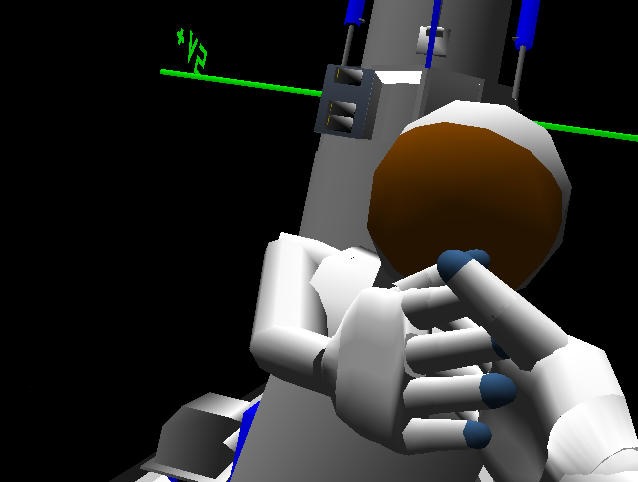} \hfill
	 \includegraphics[width=0.2\linewidth]{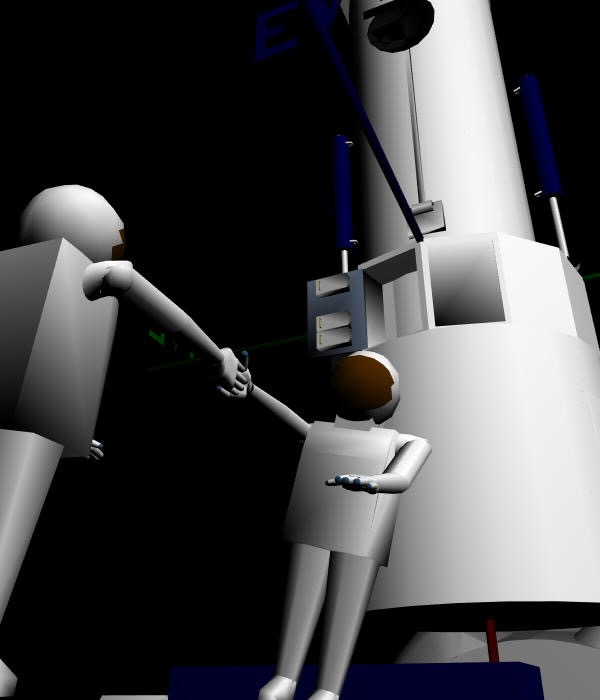}
	\caption{Scenes from the astronaut training in a distributed virtual environment
				experiment between NASA's Johnson Space Center, Houston, and Fraunhofer IGD, Damrstadt
				in 1995. From left to right: virtual environment, astronaut Ulf Merbold, 
				handshake, handshake from third person view.}
 \label{fig:astronaut}
\end{figure*}

One of the milestones at Fraunhofer IGD was the 
experiment demonstrating the feasbility of collaborative training
of distributed astronaut~\cite{Felger1996}%
\footnote{
	Incidentally, at this panel the panelists also discussed
	ideas of systems and experiences that we now call \textsl{metaverse},
	and which were dubbed ``large-scale social electronic spaces''
	at the time.
}.
The experiment was performed between NASA's Johnson Space Center (JSC), Houston,
and Fraunhofer, Darmstadt, on September 20, 1996.
Astronauts Bernard Harris (NASA) at JSC and Ulf Merbold (ESA) in Darmstadt
performed the training of the replacement of a module of the Hubble space telescope
that was, in the virtual environment, docked to the space shuttle
(see Figure~\ref{fig:astronaut}).

The virtual environment (VE) of the astronaut training experiment was comprised of 18\,000 triangles;
both astronauts used an HMD with dataglove and electromagnetic tracking.
For navigation, JSC used two joysticks while IGD used a spacemouse.
Communication of all updates in the VE was done over an ISDN connection with $2\times 64$ kbit/sec
with an update rate of 5\,Hz. Signal latency was 0.1\,sec, with occasional lags of 2 seconds.
Both sides used their own, proprietary VR software.

The experiment lasted for 30 minutes and was deemed very succesful by both astronauts.
While it was (and still is) hard to simulate zero gravity, both astronauts felt that the technology
could help reduce travel time or provide ``ad-hoc'' training of procedures for astronauts already in orbit.

Since the department had a lot of experience with scientific visualization,
we demonstrated VR as a tool for presenting numerical simulation results 
using scientific visualization techniques in an immersive and interactive
way. One example was our virtual wind tunnel (Fig.~\ref{fig:scivis}a), which was developed
in cooperation with Volkswagen. Users could switch between several visualization techniques,
such as particle flow or streamlines.
Results of Fraunhofer's sound propagation simulation, and results of our global illumination
software 
could also be presented immersively and were demonstrated at Siggraph 1994,
in the VROOM exhibition using a Cave~\cite{Siggraph1994visual}
(Fig.~\ref{fig:scivis}b).
In our VROOM demo, simulation parameters
and room properties could be modified interactively, and the resulting
optic and acoustic energy distributions could be visualized and experienced in the Cave.








\section{Conclusions}

With this article, we have outlined and highlighted some of the research done 
at Fraunhofer IGD, Damrstadt, Germany, in the years 1991 -- 2000.
This demonstrates that the Fraunhofer IGD was one of the frontrunners
of VR research in Germany and, arguably, in Europe in the 1990's.
It was instrumental in bringing VR to the German automotive industries
and in providing a German perspective on VR to the general public.


\acknowledgments{
The authors wish to thank, in no particular order,
Peter Astheimer, Fan Dai, Stefan Müller, Christian Knöpfle, Alexander Rettig,
Wolfgang Müller-Wittig, Matthias Unbescheiden, Bernd Lutz, Jakob , 
They were, in various roles and different times, among the main scientists and drivers
of the VR research at the Fraunhofer IGD department for VR in the 1990's.
}

\bibliographystyle{abbrv-doi}

\bibliography{Archiving-VR}

\end{document}